\documentclass[12pt]{iopart}

\usepackage{amsfonts, amssymb, mathrsfs, times, float, color, bm}
\usepackage{graphicx}
\usepackage{graphicx,epstopdf}
\usepackage{dcolumn}
\usepackage{bm}
\newcommand{\nn}{\nonumber}

\begin{document}

\title[Second order Moving Kerr-Newman Deflection]{Analytical Derivation of Second-Order Deflection in Equatorial Plane of a Radially Moving Kerr-Newman Black Hole}

\author{Guansheng He \& Wenbin Lin}

\address{School of Physical Science and Technology, Southwest Jiaotong University, Chengdu 610031, China}
\ead{wl@swjtu.edu.cn}
\vspace{10pt}

\begin{abstract}
In this work, we base on the second-order post-Minkowskian equations of motion, and apply an iterative technique to analytically derive the gravitational deflection of the relativistic particles in the equatorial plane of Kerr-Newman black hole with a radial (or longitudinal) and constant velocity. We find that the kinematically correctional effects on the second-order contributions to the deflection can be expressed into a very compact form, which are valid for both the massive particle and photon. Our result reduces to the previous formulation for the first-order deflection caused by a moving Schwarzschild black hole when the second-order contributions are dropped.
\end{abstract}


\vspace{1pc}

\section{Introduction}
Gravitational lensing serves as one of the most powerful tools in modern astrophysics. In the last two decades, the kinematically correctional effects~\cite{Heyrovsky2005,PB1993} induced by the lens' motion on the gravitational deflection have attracted considerable attentions of the relativity community
~\cite{PB1993,KopeiSch1999,KopeiMash2002,FKN2002,Frittelli2003a,Frittelli2003b,MB2003,WS2004,Sereno2005,KopeiFoma2007,KM2007}. Previous works were mainly concentrated on the velocity effects in the first-order gravitational deflection of test particles. Especially, Wucknitz and Sperhake~\cite{WS2004} investigated the deflection of test particles in the time-dependent field of a radially (or longitudinally) moving Schwarzschild source in the first post-Minkowskian (PM) approximation. Given the rapid progresses achieved in astronomical observations~\cite{Perryman2001,Gouda2004,Laskin2006,Lindegren2007,SN2009,Malbet2012}, we can expect that the kinematical effects on the higher-order deflection might be detectable in near future and are therefore worthy of our considerations. Recently, on the basis of numerical calculations, we studied the velocity-correctional effects on the second-order contributions to the gravitational deflection of relativistic test particles including photon in the equatorial plane of a radially and constantly moving Kerr-Newman (KN) black hole~\cite{LinHe2016}. In that work, we have obtained the analytical form for the second-order deflection angle of a photon by the moving black hole. However, a general formulation of the second-order deflection for both relativistic massive particles and photon due to the moving lens has not been achieved.

In this paper, we start with the 2PM equations of motion of test particles, and then apply an iterative technique to analytically calculate the second-order gravitational deflection of relativistic test particles including photon in the equatorial plane of a moving KN black hole with a radial constant velocity. We constrain our discussions in the small-angle, weak-field, and thin-lens approximation. This paper is organized as follows. In Section~\ref{derivation}, we first review the 2PM equations of motion for test particles, and then derive the second-order deflection angle of the test particles. Section~\ref{Discussions} presents the discussions of the formulation and a summary.

Natural units in which $G = c = 1$ are used throughout the paper.

\section{Second-order gravitational deflection caused by the moving Kerr-Newman source} \label{derivation}

\subsection{Equations of motion up to second post-Minkowskian order for test particles}
The 2PM equations of motion for test particles in the field of a longitudinally and constantly moving KN black hole can be calculated directly from the corresponding metric~\cite{LinHe2015}, which can be obtained from the harmonic KN metric~\cite{LinJiang2014} via a Lorentz boosting. Let $\{\bm{e}_1,~\bm{e}_2,~\bm{e}_3\}$ denote the orthonormal basis of a three-dimensional Cartesian coordinate system. We assume the rest frame of the background spacetime and the comoving frame for the barycentre of the gravitational source to be $(t,~x,~y,~z)$ and $(X_0,~X_1,~X_2,~X_3)$, respectively. The 2PM equations of motion for test particles propagating in the equatorial plane of a moving KN source with a constant radial velocity $\bm{v}=v\bm{e}_1$ can be written as follows~\cite{LinHe2016}
{\small \begin{eqnarray}
\hspace*{-70pt}\nn0=\ddot{t}+\frac{v\,\gamma^3\,\dot{t}^{\hspace*{1pt} 2}\,X_1}{R^2}\!\left[-\frac{(1+v^2)\,M}{R}+\frac{(v^2-4)\,M^2+Q^2}{R^2}
+\frac{v^2(M^2\!-\!Q^2)\,(y^2-X^2_1)}{R^4}+\frac{6\,v\,aMy}{R^3}\right]  \\
\hspace*{-70pt}\nn+\frac{\gamma^3\,\dot{x}^2X_1}{R^2}\!\left\{\frac{v\,(v^2\!-\!3)\,M}{R}+\frac{v\,[\,(1\!-\!4v^2)\,M^2\!+\!(\,2\!-\!v^2\,)\,Q^2\,]}{R^2}
\!+\!\frac{v\,(M^2\!-\!Q^2)\,(y^2\!-\!X^2_1)}{R^4}\!+\!\frac{6\,a\,M\,y}{R^3}\right\}  \\
\hspace*{-70pt}\nn+\frac{2\gamma^3\,\dot{t}\,\dot{x}X_1}{R^2}\!\!\left[\frac{(1\!+\!v^2)M}{R}\!+\!\frac{3v^2M^2\!-\!Q^2}{R^2}
\!-\!\frac{v^2(M^2\!-\!Q^2)(y^2\!-\!X^2_1)}{R^4}\!-\!\frac{6\,v a M y}{R^3}\right]\!\!+\!\frac{2(1\!+\!v^2)\gamma^2M\dot{t}\,\dot{y}\,y}{R^3}  \\
\hspace*{-70pt}-\frac{4\,v\gamma^2\,M\,\dot{x}\,\dot{y}\,y}{R^3}~,\label{EM1}
\end{eqnarray}
\begin{eqnarray}
\hspace*{-70pt}\nn 0=\ddot{x}\!+\!\frac{\gamma^3\hspace*{1.5pt}\dot{t}^{\hspace*{1pt} 2}X_1}{R^2}\!\left[\frac{(1\!-\!3v^2)M}{R}\!+\!\frac{(v^2\!-\!4)M^2\!+\!(2v^2\!-\!1)Q^2}{R^2}
+\frac{v^2(M^2\!-\!Q^2)\,(y^2\!-\!X^2_1)}{R^4}\!+\!\frac{6v^3aMy}{R^3} \right] \\
\hspace*{-70pt}\nn +\frac{\gamma^3\hspace*{1.5pt}\dot{x}^2X_1}{R^2}\!\left[-\frac{(1\!+\!v^2)M}{R}\!+\!\frac{(1\!-\!4v^2)M^2\!+\!v^2Q^2}{R^2}
\!+\!\frac{(M^2\!-\!Q^2)\,(y^2\!-\!X^2_1)}{R^4}\!+\!\frac{6vaMy}{R^3}\right]\!+\!\frac{2\,v\hspace*{1pt}\gamma^3\,\dot{t}\,\dot{x}X_1}{R^2} \\
\hspace*{-70pt} \times\!\left[\frac{(1\!+\!v^2)\,M}{R}\!+\!\frac{3M^2\!-\!v^2\,Q^2}{R^2}\!-\!\frac{(M^2\!-\!Q^2)\,(\hspace*{1.5pt}y^2\!-\!X^2_1\hspace*{1.5pt})}{R^4}
\!-\!\frac{6\,v\,aMy}{R^3}\right]+\frac{2\,\gamma\,M\,(v\dot{X}_0\!-\!\dot{X}_1)\,\dot{y}\,y}{R^3} ~,~ \label{EM2}
\end{eqnarray}
\begin{eqnarray}
\hspace*{-70pt}\nn 0=\ddot{y}+\dot{t}^{\hspace*{1pt} 2}
\left\{\frac{\gamma^2\,\,y}{R^2}\left[\frac{(1+v^2\,)\,M}{R}\!-\!\frac{(4+v^2)\,M^2+Q^2}{R^2}\!+\!\frac{v^2\,(M^2\!-\!Q^2\,)\,\,(y^2\!-\!X^2_1)}{R^4}\right]
\!-\!\frac{2\,v\,\gamma^2\,aM}{R^3}\right\}  \\
\hspace*{-70pt}\nn +\,\,\dot{x}^2\left\{\frac{\gamma^2\,\,y}{R^2}\left[\,\frac{(\,1+v^2\,)\,M}{R}\!-\!\frac{(\,1+4\,v^2\,)\,M^2+v^2\,Q^2}{R^2}
\!+\!\frac{(\,M^2\hspace*{-1.2pt}-\hspace*{-0.8pt}Q^2\,)\,\,(\,y^2\!-\!X^2_1\,)}{R^4}\,\right]\!-\!\frac{2\,v\,\gamma^2\,aM}{R^3}\,\right\}  \\
\hspace*{-70pt} +\, 2\,\gamma^2\,\dot{t}\,\dot{x}\!\left[-\frac{2vMy}{R^3}\!+\!\frac{v(5M^2\!+Q^2)y}{R^4}
\!-\!\frac{v(M^2\!-\!Q^2)(y^2\!-\!X^2_1)y}{R^6}\!+\!\frac{(1\!+\!v^2)aM}{R^3}\right]\!-\!\frac{2M\dot{X}_1\dot{y}X_1}{R^3}~. \label{EM3}
\end{eqnarray}}
Here, $\gamma=(1-v^2)^{-\scriptstyle \frac{1}{2}}$ is the Lorentz factor and a dot denotes derivative with respect to the affine parameter $\xi$ which describes the trajectory~\cite{WS2004,Weinberg1972}. $M$, $Q$, and $\bm{J} (=\!J\bm{e}_3)$ are the rest mass, electrical charge, and angular momentum vector (along positive $X_3$ axis) of the gravitational source, respectively. $\Phi\equiv-\frac{M}{R}=-\frac{M}{\sqrt{X_1^2+X_2^2-a^2}}$ denotes the Newtonian gravitational potential, with $a \!=\! \frac{J}{M}$ being the angular momentum per mass. The relation $M^2\geq a^2+Q^2$ is assumed to avoid naked singularity for the black hole. Notice that Eqs.~(\ref{EM1}) - (\ref{EM3}) correspond to the $t,~x,$ and $y\,-$component of the geodesic equations in the background's rest frame, respectively, and that the residual component is trivial and omitted for the equatorial-plane propagation ($z=\frac{\partial }{\partial z}=0$). The coordinates $X_0,~X_1,~X_2$, and $X_3$ are related to $t,~x,~y$, and $z$ by the Lorentz transformation
\begin{eqnarray}
\hspace*{-50pt} X_0\equiv T=\gamma(t-vx)~, ~~~~ X_1\equiv X=\gamma(x-vt)~, ~~~~ X_2\equiv Y=y~, ~~~~ X_3\equiv Z=z~.            \label{LT}
\end{eqnarray}

\begin{figure*}[t]
\begin{center}
  \includegraphics[width=14.5cm]{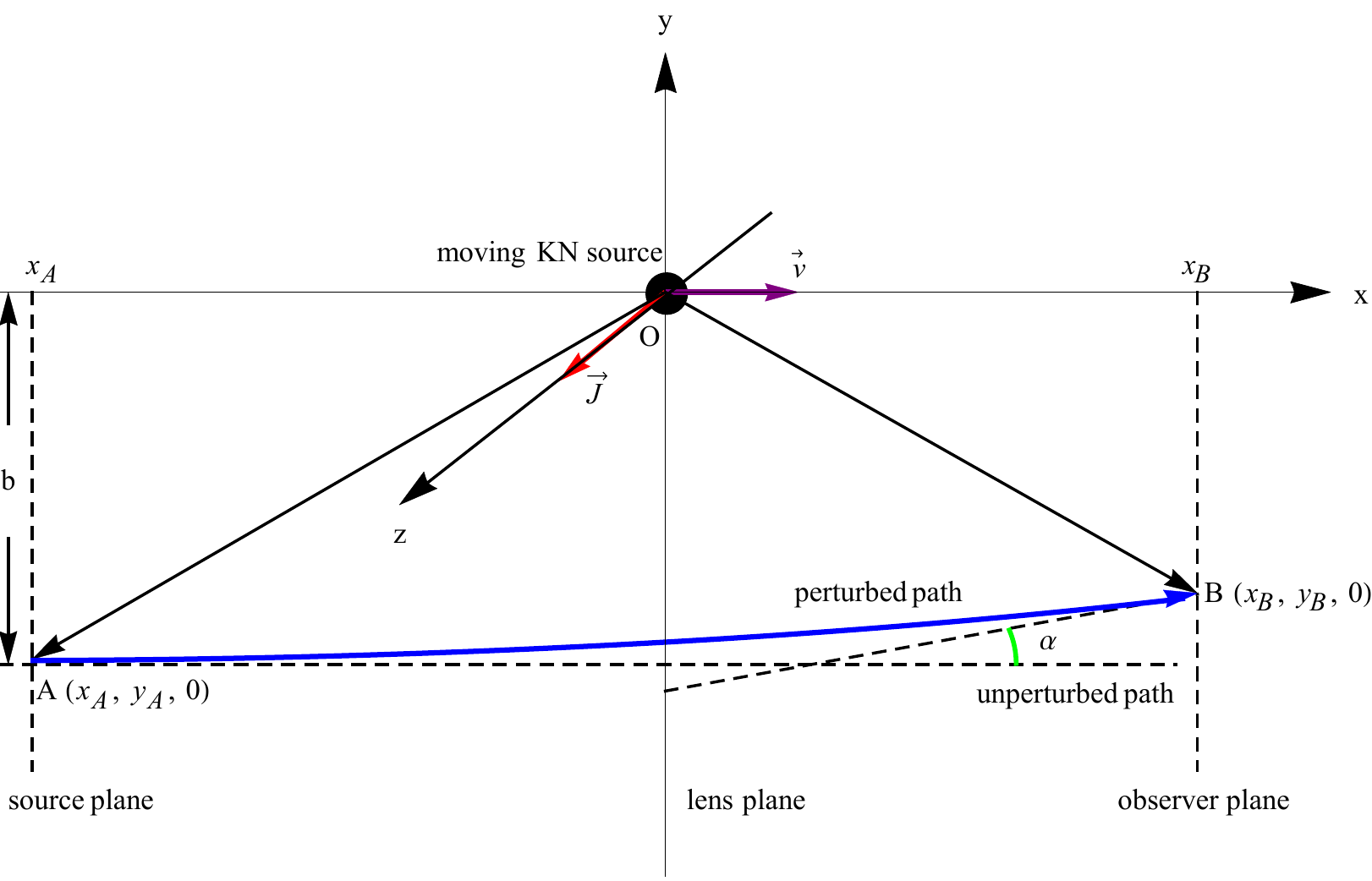}
  \caption{Schematic diagram for the gravitational deflection of a test particle caused by a longitudinally moving Kerr-Newman black hole. We assume the test particle takes the prograde motion relative to the rotation of the gravitational source.  }    \label{Figure1}
\end{center}
\end{figure*}
The schematic diagram for the deflection of test particles caused by a longitudinally moving KN source is shown in Fig.~\ref{Figure1}. The spatial coordinates of the test-particle source (denoted by $A$ on the source plane) and observer (denoted by $B$ on the observer plane) are assumed as $(x_A,~y_A,~0)$ and $(x_B,~y_B,~0)$ respectively in the background's rest frame, where $y_A\approx -b$, $x_A\ll-b$, and $x_B\gg b$, with $b$ being the impact parameter. In the comoving frame, the locations of $A$ and $B$ are denoted by $(X_A,~Y_A,~0)$ and $(X_B,~Y_B,~0)$, respectively. The solid blue line represents the perturbed propagation path of a test particle coming from $x=-\infty$ with a relativistic initial velocity $\bm{w}|_{x\rightarrow-\infty}~(\approx \bm{w}|_{x\rightarrow x_A})~=w\bm{e}_1~$ ($0<w\leq1$). We assume $v<w$~\cite{WS2004,LinHe2016} and that the test particle can pass by the black hole to reach the observer. The gravitational deflection is greatly exaggerated to distinguish between the perturbed and unperturbed (dashed horizontal line) paths.

\subsection{Second-order deflection angle due to the moving KN black hole} \label{movingKNangle}
The gravitational deflection angle for a test particle traveling from $A$ to $B$ is defined as $\alpha\equiv  \arctan\left. \frac{dy}{dx}\right|_{B}-\arctan \left.\frac{dy}{dx}\right|_{A}$. In the following, we employ an iterative technique to derive the gravitational deflection angle up to the second order, which can be written as
\begin{equation}
\alpha = \left. \frac{dy}{dx}\right|_{B}- \left.\frac{dy}{dx}\right|_{A} = \left. \frac{\dot{y}}{\dot{x}}\right|_{B}-\left. \frac{\dot{y}}{\dot{x}}\right|_{A}~.\label{alpha}
\end{equation}

Up to the zero order, Eqs.~(\ref{EM1}) - (\ref{EM3}) result in
\begin{eqnarray}
&&\dot{t}=\frac{1}{w}+O(M)~, \label{ZO-dot-t} \\
&&\dot{x}=1+O(M)~, \label{ZO-dot-x} \\
&&\dot{y}=0+O(M)~, \label{ZO-dot-y}
\end{eqnarray}
where we have followed Wucknitz and Sperhake's idea~\cite{WS2004} and used the boundary conditions $\dot{t}|_{\xi\rightarrow -\infty}=\dot{t}|_{x\rightarrow -\infty}=\frac{1}{w}$, $\dot{x}|_{\xi\rightarrow -\infty}=\dot{x}|_{x\rightarrow -\infty}=1$, and $\dot{y}|_{\xi\rightarrow -\infty}=\dot{y}|_{x\rightarrow -\infty}=0$. From Eqs.~(\ref{ZO-dot-x}) - (\ref{ZO-dot-y}), we can get a zero-order parameter transformation and the zero-order value for $y$
\begin{eqnarray}
&&dx=\left[1+O(M)\right]\,d\xi~, \label{ZO-PT-dx/dxi} \\
&&y=-b\,\left[1+O(M)\right]~,~~ \label{ZO-y}
\end{eqnarray}
where the boundary condition $y|_{\xi\rightarrow -\infty}=y|_{x\rightarrow -\infty}=-b$ has been adopted.

Substituting Eqs.~(\ref{ZO-dot-t}) - (\ref{ZO-dot-y}) and (\ref{ZO-y}) into Eqs.~(\ref{EM1}) - (\ref{EM3}) and integrating the latter equations over $\xi$, we can obtain the analytical first-order forms for $\dot{t}$, $\dot{x}$, and $\dot{y}$ as follows:
\begin{eqnarray}
&&\dot{t}=\frac{1}{w}-\frac{\gamma^2\left[(1+v^2)\left(\frac{v}{w^2}-\frac{2}{w}\right)+v(3-v^2)\right]}{1-\frac{v}{w}}
\frac{M}{\sqrt{X_1^2+b^2}}+O(M^2)~, \label{FO-dot-t} \\
&&\dot{x}=1-\frac{\gamma^2\left[(1+v^2)\left(1-\frac{2v}{w}\right)-\frac{1-3v^2}{w^2}\right]}{1-\frac{v}{w}}\frac{M}{\sqrt{X_1^2+b^2}}+O(M^2)~, \label{FO-dot-x} \\
&&\dot{y}=\frac{\gamma\left[(1+v^2)\left(1+\frac{1}{w^2}\right)-\frac{4v}{w}\right]}{1-\frac{v}{w}}\frac{M}{b}\left(1+\frac{X_1}{\sqrt{X_1^2+b^2}}\right)+O(M^2)~,  \label{FO-dot-y}
\end{eqnarray}
where Eq.~(\ref{ZO-PT-dx/dxi}) and the zero-order parameter transformation $dX_1\!=\!\left[\gamma\left(1\!-\!\frac{v}{w}\right)+O(M)\right]dx$~\cite{WS2004} have been used, and the integral constants have been determined in the lens's comoving frame. With the help of Eqs.~(\ref{ZO-PT-dx/dxi}) and (\ref{ZO-y}), we integrate Eq.~(\ref{FO-dot-y}) over $\xi$ and obtain the explicit form for $y$ up to the 1PM order as
\begin{eqnarray}
y=-b\left\{\!1\!-\!\frac{\left[(1\!+\!v^2)\left(1\!+\!\frac{1}{w^2}\right)\!-\!\frac{4v}{w}\right]}{\left(1-\frac{v}{w}\right)^2}
\frac{M\!\left(\!\sqrt{\!X_1^2\!+\!b^2}\!+\!X_1\!\right)}{b^2}+O(M^2)\!\right\}
~.~~~~~~~~\label{FO-y}
\end{eqnarray}

We then plug Eqs.~(\ref{FO-dot-t}) - (\ref{FO-y}) into Eq.~(\ref{EM3}), ignore the third and higher order terms, and have
{\small \begin{eqnarray}
\hspace*{-70pt}\nn&&\ddot{y}=\frac{\gamma^2\left[(1\!+\!v^2)\left(1\!+\!\frac{1}{w^2}\right)\!-\!\frac{4\,v}{w}\right]M\,b}{\left(X_1^2+b^2\right)^{\frac{3}{2}}}
\!+\!\frac{2\,\gamma^2\!\left[\hspace*{1pt}\frac{3}{w^2}-1-\frac{3\,v}{w}\,(1\!+\!w^2)\left(\frac{1}{w^2}\!-\!\frac{v}{w}\right)
+\frac{v^3}{w^3}\,(1\!-\!3\,w^2\hspace*{1pt})\hspace*{1pt}\right]\!M^2\hspace*{1.7pt}b}{\left(1-\frac{v}{w}\right)\left(X_1^2+b^2\right)^2}  \\
\hspace*{-70pt}\nn&&\hspace*{10pt}-\!\frac{\gamma^2\!\left[(1\!+\!v^2)\!\left(1\!+\!\frac{1}{w^2}\right)\!-\!\frac{4v}{w}\right]^2\!M^2\!
\left(\!\!\sqrt{\!X_1^2\!+\!b^2}\!+\!X_1\!\right)\!\left(X_1^2\!-\!2b^2\right)}{\left(1-\frac{v}{w}\right)^2\left(X_1^2+b^2\right)^{\frac{5}{2}}b}
\!+\!\frac{2\gamma^2\!\left[(1\!+\!v^2)\!\left(1\!+\!\frac{1}{w^2}\right)\!-\!\frac{4v}{w}\right]\!M^2\!X_1}{b\left(X_1^2+b^2\right)^{\frac{3}{2}}} \\
\hspace*{-70pt}\nn&&\hspace*{10pt}\times \! \left(\!1\!+\!\frac{X_1}{\sqrt{X_1^2+b^2}}\!\right)\!-\!\frac{\gamma^2\,b}{\left(X_1^2\!+\!b^2\right)^2}\!
\left[\hspace*{1pt}(\hspace*{1pt}1\!+\!4\,v^2\hspace*{1pt})M^2\!+\!v^2\,Q^2\!-\!\frac{4\,v\,(\,2M^2\!+\!Q^2\,)}{w}\!+\!\frac{(4\!+\!v^2)M^2\!+\!Q^2}{w^2}\hspace*{1pt}\right] \\
\hspace*{-70pt}&&\hspace*{10pt} +\frac{\gamma^2(M^2\!-\!Q^2)
\left[\left(1\!+\!\frac{v^2}{w^2}\right)b^2\!-\!\left(1\!-\!\frac{4v}{w}\!+\!\frac{v^2}{w^2}\right)X_1^2\right] b}{\left(X_1^2+b^2\right)^3}
\!-\!\frac{2\,\gamma^2\!\left[\frac{1+v^2}{w}\!-\!v\left(1\!+\!\frac{1}{w^2}\right)\right] aM}{\left(X_1^2+b^2\right)^{\frac{3}{2}}} ~, ~ \label{dott-y}
\end{eqnarray}}
in which the zero-order relation $\dot{X}_1=\gamma\left(\dot{x}-v\dot{t}\,\right)=\gamma\left(1-\frac{v}{w}\right)$ has been used. In addition, Eq.~(\ref{FO-dot-x}) gives the first-order parameter transformation
\begin{equation}
d\xi=\left\{1+\frac{\gamma^2\left[(1+v^2)\left(1-\frac{2v}{w}\right)-\frac{1-3v^2}{w^2}\right]}{1-\frac{v}{w}}\frac{M}{\sqrt{X_1^2+b^2}}+O(M^2)\right\}dx ~.  \label{PT}
\end{equation}

Therefore, we can apply Eq.~(\ref{PT}) to the integration of Eq.~(\ref{dott-y}) over $\xi$, up to the 2PM order, and obtain
\begin{eqnarray}
\hspace*{-70pt}\nn&&\dot{y}\!=\!\!\!\int\!\!  \Bigg{\{}\!
\frac{\gamma^2\!\left[(1\!+\!v^2)\left(1\!+\!\frac{1}{w^2}\right)\!-\!\frac{4v}{w}\right]\!M b}{\left(X_1^2+b^2\right)^{\frac{3}{2}}}
\!-\!\frac{\gamma^2\!\left[(1\!+\!4v^2)M^2\!+\!v^2Q^2\!-\!\frac{4v(2M^2+Q^2)}{w}\!+\!\frac{(4+v^2)M^2+Q^2}{w^2}\right]\hspace*{-1pt}b}{\left(X_1^2+b^2\right)^2} \\
\hspace*{-70pt}\nn&&\hspace*{10pt}-\frac{\gamma^4\!\left\{\!1-\frac{6}{w^2}+\frac{4\,v\left[1+3w^2+2v^2(1\,+\,w^2)+v^4(1\,-\,w^2)\right]}{w^3}
+\frac{1-2\,v^2\,(1\,+\,6w^2\,+\,5w^4)-v^4\,(3\,+\,6w^2\,-\,5w^4)}{w^4}\!\right\}\!M^2b}{\left(1-\frac{v}{w}\right)\left(X_1^2+b^2\right)^2}  \\
\hspace*{-70pt}\nn&&\hspace*{10pt}-\hspace*{0.5pt}\frac{\gamma^2\!\left[\hspace*{1pt}(\hspace*{0.5pt}1+v^2\hspace*{0.5pt})\left(1+\frac{1}{w^2}\right)
\!-\!\frac{4\hspace*{1pt}v}{w}\hspace*{1pt}\right]^2\!M^2\left(\sqrt{X_1^2+b^2}+X_1\right)}
{\left(1-\frac{v}{w}\right)^2\left(X_1^2+b^2\right)^{\frac{3}{2}}b}+2\,\gamma^2\left[(\hspace*{0.5pt}1+v^2\hspace*{0.5pt})\left(\!1\!+\!\frac{1}{w^2}\!\right)\!-\!\frac{4v}{w}\right] \\
\hspace*{-70pt}\nn&&\hspace*{10pt}\times\!\frac{M^2X_1}{\left(X_1^2+b^2\right)^{\frac{3}{2}}b}\left(\!1\!+\!\frac{X_1}{\sqrt{X_1^2+b^2}}\!\right)
\!+\!\frac{3\,\gamma^2\!\left[\hspace*{0.5pt}(\hspace*{0.5pt}1\!+\!v^2\hspace*{0.5pt})\left(1\!+\!\frac{1}{w^2}\right)
\!-\!\frac{4\,v}{w}\hspace*{0.5pt}\right]^2\hspace*{-1.1pt}M^2\,b\left(\sqrt{X_1^2+b^2}\!+\!X_1\right)}
{\left(1-\frac{v}{w}\right)^2\left(X_1^2+b^2\right)^{\frac{5}{2}}}   \\
\hspace*{-70pt}&&\hspace*{10pt}+\!\frac{\gamma^2(M^2\!-\!Q^2)\!\left[\left(1\!+\!\frac{v^2}{w^2}\right)b^2
\!-\!\left(1\!-\!\frac{4\hspace*{0.5pt}v}{w}\!+\!\frac{v^2}{w^2}\right)\!X_1^2\right]b}{\left(X_1^2+b^2\right)^3}
\!-\!\frac{2\gamma^2\!\left[\frac{1+v^2}{w}\!-\!v\left(1\!+\!\frac{1}{w^2}\right)\right]\!a\hspace*{0.5pt}M}
{\left(X_1^2+b^2\right)^{\frac{3}{2}}} \! \Bigg{\}}dx~.~~~ \label{dot-y}
\end{eqnarray}
In order to integrate Eq.~(\ref{dot-y}) more conveniently, we perform a coordinate transformation $dX_1=\gamma(1- v\dot{t}/\dot{x})dx$ which is to be calculated up to the first post-Minkowskian order. Considering Eqs.~(\ref{FO-dot-t}) - (\ref{FO-dot-x}), we can get
\begin{eqnarray}
\hspace*{-70pt}dx=\frac{1}{\left(1\!-\!\frac{v}{w}\right)\!\gamma}\!\left\{\!1\!-\!\frac{\gamma^2\!\left[3(1\!+\!v^2)\left(\frac{v^2}{w^2}\!-\!\frac{v}{w}\right)
\!+\!v^2(3\!-\!v^2)\!+\!\frac{v\,(1-3\,v^2)}{w^3}\right]\!M}{\left(1-\frac{v}{w}\right)^2\sqrt{X_1^2+b^2}}+O(M^2)\!\right\}\!dX_1~.~~~~~~~\label{transform}
\end{eqnarray}

We then substitute Eq.~(\ref{transform}) into Eq.~(\ref{dot-y}) and have
\begin{eqnarray}
\hspace*{-70pt}\nn&&\dot{y}= \frac{\gamma\left[(\hspace*{1pt}1+v^2\hspace*{1pt})\hspace*{1.5pt}\left(1+\frac{1}{w^2}\right)-\frac{4\hspace*{0.5pt}v}{w}\right]}{1-\frac{v}{w}}
\!\!\int\!\! \frac{M\,b}{\left(X_1^2+b^2\right)^{\hspace*{1pt}\frac{3}{2}}}\,dX_1
-\frac{\gamma\left(\,\frac{1}{w^2}-\frac{4\,v}{w}+v^2\,\right)}{1-\frac{v}{w}} \!\!\int\!\! \frac{Q^2\,b}{\left(X_1^2+b^2\right)^{\hspace*{1pt}2}}\,dX_1   \\
\hspace*{-70pt}\nn&&\hspace*{11pt}-\hspace*{1pt}\frac{\gamma\left[\,2\,\left(1-\frac{v}{w}-\frac{1}{w^2}-\frac{v^3}{w^3}\right)
\,+\,\frac{1\,-\,2\,v^2\,\hspace*{0.5pt}(\,1\,-\,3\,w^2\,+\,w^4\,)\,+\,v^4\,\hspace*{0.5pt}(\,2\,-\,2\,w^2\,+\,w^4\,)}{w^4}\right]}{\left(1-\frac{v}{w}\right)^3}
\!\!\int\!\! \frac{M^2\,b}{\left(\hspace*{1pt}X_1^2\,+\,b^2\hspace*{1pt}\right)^{\hspace*{1pt}2}}\,dX_1    \\
\hspace*{-70pt}\nn&&\hspace*{11pt}-\frac{\gamma\!\left[(\hspace*{0.5pt}1\!+\!v^2\hspace*{0.5pt})\left(1\!+\!\frac{1}{w^2}\right)\!-\!\frac{4\hspace*{1pt}v}{w}\right]^2}
{\left(1-\frac{v}{w}\right)^3}\!\!\int\!\frac{M^2\!\left(\!\sqrt{\!X_1^2\!+\!b^2}\!+\!X_1\!\right)}{\left(X_1^2+b^2\right)^{\frac{3}{2}}b}dX_1
\!+\!\frac{2\gamma}{1\!-\!\frac{v}{w}}\!\!\left[(\hspace*{1pt}1\!+\!v^2\hspace*{1pt})\!\left(\!1\!+\!\frac{1}{w^2}\!\right)\!-\!\frac{4v}{w}\right]        \\
\hspace*{-70pt}\nn&&\hspace*{11pt}\times \!\!\int\!\! \frac{M^2X_1}{\left(\,X_1^2+b^2\,\right)^{\,\frac{3}{2}}b}\left(\!1+\frac{X_1}{\sqrt{X_1^2+b^2}}\!\right)dX_1
\hspace*{1.1pt}+\hspace*{1.1pt}\frac{3\,\gamma}{\left(1-\frac{v}{w}\right)^3}\left[\,(\,1+v^2\,)\,\left(\,1+\frac{1}{w^2}\,\right)-\frac{4\,v}{w}\,\right]^2   \\
\hspace*{-70pt}\nn&& \hspace*{11pt}\times\!\!\!\int\!\frac{M^2b\left(\!\sqrt{\!X_1^2\!+\!b^2}\!+\!X_1\!\right)}{\left(X_1^2+b^2\right)^{\frac{5}{2}}}dX_{\!1}
\!+\!\frac{\gamma}{1\!-\!\frac{v}{w}}\!\int\!\frac{\left(M^2\!-\!Q^2\right)b\!\left[\left(\hspace*{-1pt}1\!+\!\frac{v^2}{w^2}\hspace*{-1pt}\right)\!b^2
\!-\!\left(\hspace*{-1pt}1\!-\!\frac{4v}{w}\!+\!\frac{v^2}{w^2}\hspace*{-1pt}\right)\!X_1^2\right]}{\left(X_1^2+b^2\right)^3}dX_1   \\
\hspace*{-70pt}\nn&&\hspace*{11pt} -\frac{2\left(1-v\,w\right)\gamma}{w}\!\int\! \frac{aM}{\left(X_1^2+b^2\right)^{\frac{3}{2}}}\,dX_1 \\
\hspace*{-70pt}\nn&&\hspace*{3pt}=\frac{\gamma\left[\,(\,1+v^2\,)\left(1+\frac{1}{w^2}\right)-\frac{4\,v}{w}\,\right]M X_1}{\left(1-\frac{v}{w}\right)b\sqrt{X_1^2+b^2}}
-\frac{\gamma\left(\,\frac{1}{w^2}-\frac{4\,v}{w}+v^2\,\right)\,Q^2}{2\left(1-\frac{v}{w}\right)b^2}\left(\frac{b\,X_1}{X_1^2+b^2}+\arctan\,\frac{X_1}{b}\hspace*{-1pt}\right)    \\
\hspace*{-70pt}\nn&&\hspace*{11pt}-\!\frac{\gamma\!\left[2\left(1\!-\!\frac{1}{w^2}\!-\!\frac{v}{w}\!-\!\frac{v^3}{w^3}\right)
\!+\!\frac{1-2\,v^2\,(1\,-\,3\,w^2\,+\,w^4)+v^4\,(2\,-\,2\,w^2\,+\,w^4)}{w^4}\right]\!M^2}
{2\left(1-\frac{v}{w}\right)^3b^2}\left(\frac{b\,X_1}{X_1^2\!+\!b^2}\!+\!\arctan\frac{X_1}{b}\hspace*{-1pt}\right)   \\
\hspace*{-70pt}\nn&&\hspace*{11pt}-\,\frac{2\,\gamma\,M^2}{\left(1-\frac{v}{w}\right)b}\left[\,(\,1+v^2\,)\left(1+\frac{1}{w^2}\right)-\frac{4\,v}{w}\,\right]\!
\left[\hspace*{1pt}\frac{X_1}{2\,\left(\,X_1^2+b^2\,\right)}+\frac{1}{\sqrt{X_1^2+b^2}}-\frac{\arctan\hspace*{1.5pt}\frac{X_1}{b}}{2\,b}\hspace*{1pt}\right]  \\
\hspace*{-70pt}\nn&&\hspace*{11pt}+\hspace*{2.5pt}\frac{3\,\gamma\left[\,(\,1+v^2\,)\,\left(1+\frac{1}{w^2}\right)-\frac{4\,v}{w}\,\right]^2M^2\,b}{\left(1-\frac{v}{w}\right)^3}\!
\left[\frac{X_1}{2\,b^2\left(X_1^2+b^2\right)}-\frac{1}{3\left(X_1^2+b^2\right)^{\frac{3}{2}}}
+\frac{\arctan\hspace*{1pt}\frac{X_1}{b}}{2\,b^3}\right]  \\
\hspace*{-70pt}\nn&&\hspace*{11pt}+\frac{\gamma\,(M^2\!-\!Q^2)\,b}{4\,w^2\,\left(1-\frac{v}{w}\right)}\!
\left[\frac{2\,(v\!-\!w)^{\hspace*{1pt}2}X_1}{\left(X_1^2+b^2\right)^2}\!+\!\frac{(v+w)^2}{b^2}\!\left(\!\frac{X_1}{X_1^2\!+\!b^2}\!+\!\frac{\arctan\frac{X_1}{b}}{b}\!\right)\!\right]
\!-\!\frac{2(1\!-\!vw)\gamma\,a M X_1}{w\, b^2\sqrt{X_1^2+b^2}}  \\
\hspace*{-70pt}&&\hspace*{11pt}+\frac{\gamma\left[(1\!+\!v^2)\!\left(1\!+\!\frac{1}{w^2}\right)\!-\!\frac{4v}{w}\right]^2\!M^2}{\left(1-\frac{v}{w}\right)^3b}\!
\left(\!\frac{1}{\sqrt{\!X_1^2\!+\!b^2}}\!-\!\frac{\arctan\frac{X_1}{b}}{b}\!\right)+C~,~~~~ \label{dot-y-2}
\end{eqnarray}
where $C$ denotes the integral constant and we have dropped the third and higher order terms in the derivation.

Finally, substituting Eqs.~(\ref{FO-dot-x}) and (\ref{dot-y-2}) into Eq.~(\ref{alpha}) and considering the conditions $X_A\ll -b$ and $X_B\gg b$, we can obtain
the 2PM gravitational deflection angle of a test particle propagating from $A~(X_A,~Y_A,~0)$ to $B~(X_B,~Y_B,~0)$ as follow
\begin{eqnarray}
\hspace*{-40pt} \alpha = N_1\frac{2\left(1+\frac{1}{w^2}\right)\!M}{b}+N_2\frac{3\left(1+\frac{4}{w^2}\right)\pi M^2}{4b^2}-N_3\frac{4aM}{wb^2}
-N_4\frac{\left(1+\frac{2}{w^2}\right)\pi Q^2}{4b^2}~,  \label{MovingKN-angle}
\end{eqnarray}
where the kinematical correctional coefficients $N_i~(i=1,~2,~3,~4)$ are the functions of $v$ and $w$:
\begin{eqnarray}
&& N_1= \frac{\left(1+v^2-\frac{4vw}{1+w^2}\right)\gamma}{1-\frac{v}{w}} ~,          \label{N1} \\
&& N_2= \frac{\left(1+\frac{v^2-10vw+4v^2w^2}{4+w^2}\right)\gamma}{1-\frac{v}{w}} ~, \label{N2} \\
&& N_3= (1-vw)\gamma ~,                                                              \label{N3} \\
&& N_4= \frac{\left(1+\frac{v^2-6vw+2v^2w^2}{2+w^2}\right)\gamma}{1-\frac{v}{w}} ~.  \label{N4}
\end{eqnarray}
Notice that in Eq.~(\ref{MovingKN-angle}) the second-order terms with the factor $\frac{1}{X_A}$ or $\frac{1}{X_B}$ have been ignored since they are much smaller than the second-order terms with the factor $\frac{1}{b^2}$ such as the term $\frac{3N_2\left(1+\frac{4}{w^2}\right)\pi M^2}{4b^2}$. The gravitational deflection angle $\alpha$ is a function of $v$ and $w$, therefore, we denote it as $\alpha(v, w)$ from now on.

In the above derivations, we assume that the trajectory parameter $\xi$ has the dimension of length~\cite{WS2004}. In fact, we can also assume it to have the dimension of time~\cite{Weinberg1972}, and the same result can be achieved. The detailed derivations are given in the Appendix.

\section{Discussion and Summary} \label{Discussions}
When the second-order contributions are dropped in Eq.~(\ref{MovingKN-angle}), we can get the first-order deflection angle induced by a radially moving Schwarzschild source~\cite{WS2004}
\begin{eqnarray}
\alpha(v, w)=\frac{2\,\gamma\left[(1+v^2)\left(1+\frac{1}{w^2}\right)-\frac{4v}{w}\right] M}{\left(1-\frac{v}{w}\right)b}~.  \label{MovingSchwar-angle}
\end{eqnarray}

Among the second-order contributions in Eq.~(\ref{MovingKN-angle}), the third term, which corresponds to the rotation-induced effect, is the same as our previous result~\cite{LinHe2014} achieved via the Euler-Lagrange method~\cite{WS2004}. The second term (the second-order moving Schwarzschild contribution) and the forth term (the charge-induced contribution) are analytically derived for the first time and completely new.

For the case of no translational motion $(v=0)$ of the gravitational source, the four kinematical coefficients reduce to $N_1=N_2=N_3=N_4=1$,
and Eq.~(\ref{MovingKN-angle}) becomes the second-order Kerr-Newman deflection angle of a massive particle $(w<1)$~\cite{LinHe2016,LinHe2017}
\begin{eqnarray}
\hspace*{-40pt}\alpha(0, w)=2\left(1+\frac{1}{w^2}\right)\frac{M}{b}+3\left(\frac{1}{4}+\frac{1}{w^2}\right)\frac{\pi M^2}{b^2}-\frac{4aM}{wb^2}
-\left(1+\frac{2}{w^2}\right)\frac{\pi Q^2}{4b^2}~,  \label{KN-massive}
\end{eqnarray}
which can be reduced to the second-order Schwarzschild deflection of massive particles~\cite{AR2002,AR2003} when we drop the rotation and electrical charge of the black hole ($a=Q=0$).

For a photon $(w=1)$, $N_i=(1-v)\gamma$, and Eq.~(\ref{MovingKN-angle}) reduces to the second-order deflection angle caused by a radially moving KN source
\begin{eqnarray}
\alpha(v, 1)=(1-v)\gamma \left(\frac{4M}{b}+\frac{15\pi M^2}{4b^2}-\frac{4aM}{b^2}-\frac{3\pi Q^2}{4b^2}\right)~,  \label{movingKN-light}
\end{eqnarray}
which is just the result based on the numerical calculations~\cite{LinHe2016}. Notice that the second and fourth terms in Eq.~(\ref{movingKN-light}) have been correctly anticipated by the numerical analysis~\cite{LinHe2016}, but they are rigorously derived here for the first time.

In the limit $v\rightarrow0$ and $w\rightarrow1$, Eq.~(\ref{MovingKN-angle}) is simplified to the second-order Kerr-Newman deflection angle of a photon~\cite{CS2015}
\begin{eqnarray}
\alpha(0, 1)=\frac{4M}{b}+\frac{15\pi M^2}{4b^2}-\frac{4aM}{b^2}-\frac{3\pi Q^2}{4b^2}~.  \label{KN-light}
\end{eqnarray}

Finally, it should be mentioned that the measurability of the kinematically correctional effects on the second-order contributions to the gravitational deflection has been discussed in our previous work~\cite{LinHe2016}.

In summary, based on the equations of motion of test particles and the iterative technique, we achieve a general formulation for the second-order gravitational deflection of the test particles in the equatorial plane of a longitudinally and constantly moving Kerr-Newman black hole. Our formulation is not only valid for photons but also for relativistic massive particles, which generalizes the results presented in the previous literatures and might be useful in the future observations.

\section*{ACKNOWLEDGEMENT}
We are grateful to the reviewers for their useful suggestions and comments on improving the quality of this paper. This work was supported in part by the National Natural Science Foundation of China (Grant Nos. 11647314 and 11547311).

\appendix

\section{Derivations of the gravitational deflection angle based on the trajectory parameter having the dimension of time} \label{Appendix}
As done in Ref.~\cite{Weinberg1972}, we now assume the trajectory parameter $\xi$ of a test particle to have the dimension of time in the following computations for the deflection angle $\alpha(v, w)$.

Considering the boundary conditions $\dot{t}|_{x\rightarrow -\infty}=1$, $\dot{x}|_{x\rightarrow -\infty}=w$, and $\dot{y}|_{x\rightarrow -\infty}=0$, we can obtain the zero-order values for $\dot{t}$, $\dot{x}$, and $\dot{y}$ from Eqs.~(\ref{EM1}) - (\ref{EM3}) as follow
\begin{eqnarray}
&&\dot{t}=1+O(M)~, \label{A1} \\
&&\dot{x}=w+O(M)~, \label{A2} \\
&&\dot{y}=0+O(M)~. \label{A3}
\end{eqnarray}
Eqs.~(\ref{A2}) - (\ref{A3}) lead to
\begin{eqnarray}
&&y=-b\left[1+O(M)\right]~, \label{A4} \\
&&d\xi=\left[\frac{1}{w}+O(M)\right]dx~, \label{A5}
\end{eqnarray}
where the boundary condition $y|_{x\rightarrow -\infty}=-b$ has been used.

We then substitute Eqs.~(\ref{A1}) - (\ref{A4}) into Eqs.~(\ref{EM1}) - (\ref{EM3}), integrate Eqs.~(\ref{EM1}) - (\ref{EM3}) over $\xi$, and get the analytical first-order forms for $\dot{t}$, $\dot{x}$, and $\dot{y}$ as
\begin{eqnarray}
&&\dot{t}=1-\frac{\gamma^2\left[(1+v^2)\left(\frac{v}{w}-2\right)+vw(3-v^2)\right]}{1-\frac{v}{w}}\frac{M}{\sqrt{X_1^2+b^2}}+O(M^2)~, \label{A6} \\
&&\dot{x}=w+\frac{w\gamma^2\left[(1+v^2)\left(\frac{2v}{w}-1\right)+\frac{1-3v^2}{w^2}\right]}{1-\frac{v}{w}}\frac{M}{\sqrt{X_1^2+b^2}}+O(M^2)~, \label{A7} \\
&&\dot{y}=\frac{\gamma\left[\frac{(1+v^2)(1+w^2)}{w}-4v\right]}{1-\frac{v}{w}}\frac{M}{b}\left(1+\frac{X_1}{\sqrt{X_1^2+b^2}}\right)+O(M^2)~,  \label{A8}
\end{eqnarray}
where Eq.~(\ref{A5}) and the parameter transformation $dX_1=\left[\gamma\left(1-\frac{v}{w}\right)+O(M)\right]dx$ have been applied.
The integration of Eq.~(\ref{A8}) over $\xi$ makes us get
{\small\begin{eqnarray}
y=-b\left\{1\!-\!\frac{\left[(1\!+\!v^2)\left(1\!+\!\frac{1}{w^2}\right)\!-\!\frac{4v}{w}\right]}{\left(1-\frac{v}{w}\right)^2}
\frac{M\!\left(\!\sqrt{\!X_1^2\!+\!b^2}\!+\!X_1\!\right)}{b^2}+O(M^2)\right\}~. ~~~~~~~~  \label{A9}
\end{eqnarray}}

Plugging Eqs.~(\ref{A6}) - (\ref{A9}) into Eq.~(\ref{EM3}) and ignoring the third and higher order terms, we have
{\small \begin{eqnarray}
\hspace*{-70pt}\nn&&\ddot{y}=\frac{\gamma^2\left[(1\!+\!v^2)(1\!+\!w^2)\!-\!4vw\right]Mb}{\left(X_1^2+b^2\right)^{\frac{3}{2}}}
\!-\!\frac{\gamma^2\left[(4\!+\!v^2\!+\!w^2\!-\!8vw\!+\!4v^2w^2)M^2\!+\!(1\!-\!4vw\!+\!v^2w^2)Q^2\right]\!b}{\left(X_1^2+b^2\right)^2}    \\
\hspace*{-70pt}\nn&&\hspace*{19pt}+\frac{2\,\gamma^2\!\left[3-w^2\!+\!3\,v\,(1+w^2)\left(v-\!\frac{1}{w}\right)+\frac{v^3\,(1-3w^2)}{w}\right]\!M^2\,b}
{\left(1-\frac{v}{w}\right)\left(X_1^2+b^2\right)^2}
-\frac{w^2\gamma^2\!\left[\left(1+v^2\right)\left(1+\frac{1}{w^2}\right)-\frac{4\,v}{w}\right]^2}{\left(1-\frac{v}{w}\right)^2} \\
\hspace*{-70pt}\nn&&\hspace*{19pt}\times\frac{M^2\left(\sqrt{\,X_1^2+b^2}+X_1\right)}{\left(X_1^2+b^2\right)^{\frac{3}{2}}b}
+\frac{2\,\gamma^2\left[\,(\,1+v^2\,)\,\,(\,1+w^2\,)-4\,v\,w\,\right]M^2X_1}{\left(X_1^2+b^2\right)^{\frac{3}{2}}b}\left(1+\frac{X_1}{\sqrt{\,X_1^2+b^2}}\right)  \\
\hspace*{-70pt}\nn&&\hspace*{19pt}+\frac{\gamma^2\left[(v^2+w^2)\,b^2-(v^2-4vw+w^2)\,X_1^2\right](M^2-Q^2)\,b}{\left(X_1^2+b^2\right)^3}
\!+\!\frac{3\gamma^2\left[\left(1+v^2\right)\left(1+w^2\right)-4vw\right]^2}{w^2\left(1-\frac{v}{w}\right)^2} \\
\hspace*{-70pt}&&\hspace*{19pt}\times\frac{M^2\,b\left(\sqrt{\!X_1^2+b^2}+X_1\right)}{\left(X_1^2+b^2\right)^{\frac{5}{2}}}
-\frac{2\,\gamma^2\left[\,w\,(1+v^2)-v\,(1+w^2)\,\right]aM}{\left(X_1^2+b^2\right)^{\frac{3}{2}}} ~. ~~~ \label{A10}
\end{eqnarray}}

Notice that Eq.~(\ref{A7}) gives the first-order parameter transformation as follow
\begin{equation}
\hspace*{-40pt}d\xi=\frac{1}{w}\left\{1-\frac{\gamma^2\left[\left(\,\frac{2\,v}{w}-1\,\right)\,(\,1+v^2\,)
+\frac{1\,-\,3\,v^2}{w^2}\right]}{1-\frac{v}{w}}\frac{M}{\sqrt{X_1^2+b^2}}+O(M^2)\right\}dx~. \label{A11}
\end{equation}
We can thus apply Eq.~(\ref{A11}) to the integration of Eq.~(\ref{A10}) over $\xi$, up to the second order, and obtain
{\small \begin{eqnarray}
\hspace*{-70pt}\nn&&\dot{y}\!=\!\!\! \int\!\!\Bigg{\{}\!\! \frac{\gamma^2\!\left[(1\!+\!v^2)(1\!+\!w^2)\!-\!4vw\right]\!Mb}{w\left(X_1^2+b^2\right)^{\frac{3}{2}}}
\!-\!\frac{\gamma^2\!\left[(4\!+\!v^2\!+\!w^2\!-\!8vw\!+\!4v^2w^2)M^2\!\!+\!(1\!-\!4vw\!+\!v^2w^2)Q^2\right]\!b}{w\left(X_1^2+b^2\right)^2}  \\
\hspace*{-70pt}\nn&&+\frac{\gamma^2M^2\,b}{\left(w\!-\!v\right)\left(X_1^2\!+\!b^2\right)^2}\!\left\{2\!\left[3\!-\!w^2\!+\!\frac{3v(1\!+\!w^2)\,(vw\!-\!1)\!+\!v^3(1\!-\!3w^2)}{w}\right]\!
\!-\!\gamma^2\!\left[(1\!+\!v^2)(1\!+\!w^2)\!-\!4vw\right] \right. \\
\hspace*{-70pt}\nn&&\left. \times\!\left[(1\!+\!v^2)\left(\frac{2v}{w}-1\right)+\frac{1\!-\!3v^2}{w^2}\,\right]\right\}
+\frac{2\gamma^2\left[\,(1+v^2)\,(1+w^2)-4vw\,\right]M^2X_1}{w\left(X_1^2+b^2\right)^{\frac{3}{2}}b}\!\left(1\!+\!\frac{X_1}{\sqrt{X_1^2+b^2}}\right)  \\
\hspace*{-70pt}\nn&&-\frac{w\,\gamma^2\left[\,(1+v^2)\left(1+\frac{1}{w^2}\right)-\frac{4\,v}{w}\,\right]^2M^2\left(\sqrt{X_1^2+b^2}+X_1\right)}
{\left(1-\frac{v}{w}\right)^2\left(X_1^2+b^2\right)^{\frac{3}{2}}b}+\frac{3\,\gamma^2\left[\,(1+v^2)\left(1+w^2\right)-4\,v\,w\,\right]^2}{w^3\left(1-\frac{v}{w}\right)^2}   \\
\hspace*{-70pt}\nn&&\times\!\frac{M^2b\!\left(\!\!\sqrt{\!X_1^2\!+\!b^2}\!+\!X_1\!\!\right)}{\left(X_1^2+b^2\right)^{\frac{5}{2}}}
\!+\!\frac{\gamma^2\!\left[(v^2\!\!+\!w^2)b^2\!\!-\!(v^2\!\!-\!4vw\!+\!w^2)X_1^2\right]\!(M^2\!\!-\!Q^2)b}{w\left(X_1^2+b^2\right)^3}
\!-\!2\gamma^2\!\!\left[1\!+\!v^2\!\!-\!\frac{v(1\!+\!w^2)}{w}\!\right] \\
\hspace*{-70pt}&&\times\frac{aM}{\left(X_1^2+b^2\right)^{\frac{3}{2}}} \Bigg{\}}dx~.~  \label{A12}
\end{eqnarray}}

In order to get the explicit form for $\dot{y}$ conveniently, we perform a 1PM coordinate transformation $dX_1=\gamma(1- v\dot{t}/\dot{x})dx$, which can be calculated via Eqs.~(\ref{A6}) - (\ref{A7}) as follow
\begin{eqnarray}
\hspace*{-70pt}dX_1\!=\left(1\!-\!\frac{v}{w}\right)\!\gamma\!\left\{\!1\!+\!\frac{\gamma^2\!\left[3(1\!+\!v^2)\!\left(\frac{v^2}{w^2}\!-\!\frac{v}{w}\right)
\!+\!v^2(3\!-\!v^2)\!+\!\frac{v(1\!-\!3v^2)}{w^3}\right]\!M}{\left(1-\frac{v}{w}\right)^2\sqrt{X_1^2+b^2}}\!+\!O(M^2)\!\right\}\!dx~.~~~~~~~~~ \label{A13}
\end{eqnarray}
Substituting Eq.~(\ref{A13}) into Eq.~(\ref{A12}), we have
{\small \begin{eqnarray}
\hspace*{-70pt}\nn\dot{y}\,\,=\int\Bigg{\{}\hspace*{0.5pt}\frac{\gamma\,\left[\,(\,1+v^2\,)\,(\,1+w^2\,)\,-\,4\,v\,w\,\right]\,M\,b}{(w-v)\left(X_1^2+b^2\right)^{\frac{3}{2}}}
-\frac{\gamma\,\,(\,1-4\,v\,w+v^2\,w^2\,)\,\,Q^2\,b}{(w-v)\left(X_1^2+b^2\right)^2}\,-\,\frac{M^2\,b}{\left(\,X_1^2+b^2\,\right)^2}  \\
\hspace*{-70pt}\nn \times\frac{\gamma\hspace*{-0.8pt}\left[1\!-\!2(wv^3\!+\!w^2\!+\!vw^3\!\!-\!w^4)\!-\!2v^2(1\!-\!3w^2\!\!+\!w^4)\!+\!v^4(2\!-\!2w^2\!\!+\!w^4)\right] }{(w-v)^3}
\!+\! \frac{2M^2X_1\!\!\left(\!\!\sqrt{\!X_1^2\!+\!b^2}\!+\!X_1\!\right)}{\left(X_1^2+b^2\right)^2b}  \\
\hspace*{-70pt}\nn \times\,\frac{\gamma\left[\,(\,1+v^2\,)\,(\,1+w^2\,)-4\,v\,w\,\right]}{w-v}
-\frac{w^4\,\gamma\left[\,(\,1+v^2\,)\,(\,1+\frac{1}{w^2}\,)-\frac{4\,v}{w}\,\right]^{\,2}M^2\left(\sqrt{\,X_1^2+b^2}+X_1\right)}{(w-v)^3\left(X_1^2+b^2\right)^{\frac{3}{2}}b} \\
\hspace*{-70pt}\nn +\frac{3\hspace*{0.5pt}\gamma\!\left[(1+v^2)\,(1+w^2)-4vw\right]^2\!M^2\,b\left(\!\sqrt{X_1^2+b^2}\!+\!X_1\!\right)}{(w-v)^3\left(X_1^2+b^2\right)^{\frac{5}{2}}}
\!+\!\frac{\gamma\left[\,(v^2\!+\!w^2)\,b^2\!-\!(v^2\!-\!4wv\!+\!w^2)\,X_1^2\,\right]}{w-v}   \\
\hspace*{-70pt}\nn\times\,\frac{(M^2-Q^2)\,b}{\left(X_1^2+b^2\right)^3}-\frac{2\,\gamma\,[\,w\,(1+v^2)-v\,(1+w^2)\,]\,a M}{(w-v)\left(X_1^2+b^2\right)^{\frac{3}{2}}} \,\Bigg{\}}\,dX_1
\end{eqnarray}}
{\small \begin{eqnarray}
\hspace*{-70pt}\nn\hspace*{16pt}=\hspace*{3pt}\frac{\gamma\left[\,(\,1+v^2\,)\,(\,1+w^2\,)-4\,v\,w\,\right]MX_1}{(w-v)\,b\,\sqrt{X_1^2+b^2}}
\!-\!\frac{\gamma\,(1-4vw+v^2w^2)\,Q^2}{2\,(w-v)\,b^2}\left(\frac{bX_1}{X_1^2+b^2}\!+\!\arctan\!\frac{X_1}{b} \right) \\
\hspace*{-70pt}\nn-\frac{\gamma\!\left[1\!-\!2(v^3w\!+\!w^2\!+\!vw^3\!\!-\!w^4)\!-\!2v^2(1\!-\!3w^2\!\!+\!w^4)\!+\!v^4(2\!-\!2w^2\!\!+\!w^4)\right]\!M^2}{2\,(w-v)^3\,b^2}\!
\!\left(\!\frac{bX_1}{X_1^2\!+\!b^2}\!+\!\arctan\!\frac{X_1}{b} \!\right) \\
\hspace*{-70pt}\nn+\,\frac{3\,\gamma \left[\,(\,1+v^2\,)\,(\,1+w^2\,)-4\,v\,w\,\right]^{\,2}M^2\,b}{(w-v)^3}\,
\left[\,\frac{X_1}{2\,\,b^2\left(\,X_1^2+b^2\,\right)}-\frac{1}{3\,\left(\,X_1^2+b^2\,\right)^{\,\frac{3}{2}}}+\frac{\arctan\frac{X_1}{b}}{2\,b^3}\,\right]  \\
\hspace*{-70pt}\nn+\frac{w^4\gamma\left[(1\!+\!v^2)(1\!+\!\frac{1}{w^2})\!-\!\frac{4\,v}{w}\right]^2\!M^2}{(w-v)^3\,b}\!\!
\left(\!\!\frac{1}{\sqrt{\!X_1^2\!+\!b^2}}\!-\!\frac{\arctan\frac{X_1}{b}}{b}\!\!\right)\!-\!\frac{2\,\gamma\left[(1+v^2)(1+w^2)-4vw\right]M^2}{(w-v)b}  \\
\hspace*{-70pt}\nn \times\!\left[\frac{X_1}{2\left(X_1^2\!+\!b^2\right)}\!+\!\frac{1}{\sqrt{X_1^2\!+\!b^2}}\!-\!\frac{\arctan\frac{X_1}{b}}{2b}\right]
\!+\!\left[\frac{2\,(v\!-\!w)^2X_1}{\left(X_1^2\!+\!b^2\right)^2}\!+\!\frac{(v\!+\!w)^2X_1}{b^2\!\left(X_1^2\!+\!b^2\right)}\!+\!\frac{(v\!+\!w)^2\arctan\frac{X_1}{b}}{b^3}\right]   \\
\hspace*{-70pt}\times\,\frac{\gamma\,(M^2-Q^2)\,b}{4\,(w-v)}-\frac{2\,\gamma\,(1-v\,w)\,aM X_1}{b^2\sqrt{X_1^2+b^2}}+C ~,~  \label{A14}
\end{eqnarray}}
where we have ignored the third and higher order terms. Based on Eqs.~(\ref{A14}) and (\ref{A7}), we can obtain the explicit form of the 2PM deflection angle $\alpha(v,w)$ by performing a definite integral over $X_1$ from $A~(X_A,~Y_A,~0)$ to $B~(X_B,~Y_B,~0)$. Considering $X_A\ll -b$ and $X_B\gg b$ and omitting the second-order terms with the factor $\frac{1}{X_A}$ or $\frac{1}{X_B}$, we can achieve the same result as Eq.~(\ref{MovingKN-angle}).

\end{document}